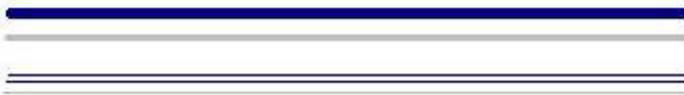
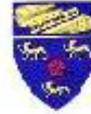

# TEORI KINETIK UNTUK EKONOMI

**wan ahmad tajuddin wan abdullah**
*jabatan fizik univeristi malaya 50603 kuala lumpur*

http://fizik.um.edu.my/hitkat?wat
wat@um.edu.my

**Abstrak**. *Mungkinkah kita bina suatu sistem ekonomi menggunakan teori ala kinetik?*

November 2002

Summary

# Kinetic Theory for Economies
*Wan Ahmad Tajuddin Wan Abdullah, Physics Department, Universiti Malaya*

A bottom-approach to econophysics modelling is taken, in the complex systems spirit, by starting with a minimal logic structure and devising a dynamics for it that is hoped to be powerful enough to produce emergent economics-like behaviour. In particular, we have $N$ agents each with three parameters: amount of money possessed, amount of (a single type of) goods possessed, and his respective assumed or asking price (for a unit of the goods), meeting each other at random, with a *buy* dynamics such that the agent with a higher assumed price buy (all that he can of) the goods from the other agent, with the former's assumed price changing to the value of the latter's, except when the latter has no more goods to sell, in which case the latter now buys from the former (all that she can) and her assumed price now increases to be the value of the former's. We simulate such a model with fixed initial capital and fixed initial particular fraction of it in the form of goods, but with initial assumed prices distributed randomly normally between the agents.

Using this model, we find an economics-like 'demand curve' where prices relax to a mean which is higher for a lower initial fraction of (lower availability of) goods. A dependence on the number of agents used is also seen, which is asking for further investigation. Also interesting is the behaviour of variances in prices between agents with respect to goods fraction, and the linear shape of the final total wealth versus goods fraction.

The economics-like demand curve is not produced when we use a converse model, with *sell* dynamics where the agent with the lower price sells to the other, with the former's price increasing to the value of the latter's, except when the latter runs out of money and sells to the former at the lower price and lowers his price to it, though the increased final prices independent of goods fraction and the linear shape of the final total capital are interesting in themselves. A symmetric model where who buys or sells does not depend on the respective agent's assumed prices also do not show such a demand curve, and is uneventful (except for a hint of increasing final total wealth with goods fraction), as expected.

The basic model is of course open to all kinds of modifications, and as a preliminary exploration, we looked at the effect of the money subjected to increase due to a fixed interest. We find different behaviour in time for two different regimes of goods fractions.

# TEORI KINETIK UNTUK EKONOMI[*]

**wan ahmad tajuddin wan abdullah**
*jabatan fizik univeristi malaya 50603 kuala lumpur*

http://fizik.um.edu.my/hitkat?wat
wat@um.edu.my

**Abstrak**. *Mungkinkah kita bina suatu sistem ekonomi menggunakan teori ala kinetik?*

## PENGENALAN

Reduktionisme mengandaikan bahawa kita boleh faham kejadian alam dengan cara memecahkan sesuatu sistem itu kepada juzuk-juzuknya dan mengkaji sifat dan telah juzuk-juzuk itu, dan seterusnya mengiferenskan sifat dan telatah sistem asal. Namun, fenomena dalam fizik menunjukkan bahawa kaedah sebegini tidak memadai, kerana apabila juzuk-juzuk suatu sistem digabungkan, walaupun dinamik juzuk-juzuk itu difahami, sifat dan telatah sistem yang terhasil kadang-kadang tidak seperti yang dijangkakan, dan ada sifat dan telatah *timbulan* hasil kerjasama dan sinergi juzuk-juzuk berkenaan. Ini mencetuskan kajian dalam sistem kompleks, dengan tujuan memahami kekompleksan yang timbul dalam gabungan unsur-unsur mudah menggunakan peralatan analisis fizik.

Sistem sosial dan sistem ekonomi boleh dilihat sebagai sistem kompleks. Pegunaan kaedah fizik dalam kajian ekonomi telah melahirkan bidang yang kini dikenali *ekonofizik*. Arah penyelidikan yang diambil dalam ekonofizik sekarang ialah kebanyakannya *atas-ke-bawah*, dengan cara membina model untuk menerangkan fenomena yang dilihat di dalam data kewangan, misalnya dalam naik-turun harga saham dan kadar pertukaran wang asing. Di sini kita ingin mempelopori suatu pendakatan *bawah-ke-atas*, dengan bermula dengan model mudah dan melihat jika apa-apa telatah keekonomian boleh timbul. Ini selaras dengan banyak pendekatan dalam sistem kompleks secara amnya (misalnya, "... *find a minimal logic structure and devise a dynamics for it that is powerful enough to simulate complex systems …*" – cari suatu struktur logik minimum dan adakan suatu dinamik untuknya yang cukup kuat untuk mensimulasikan sistem kompleks [1]).

## MODEL BELIAN

Kita anggapkan sistem kita terdiri daripada $N$ agen $i$ yang setiapnya mempunyai 3 parameter: $d_i$ yang mewakili amaun 'duit' yang dimilikinya, $b_i$ yang mewakili amaun 'barangan' yang dimilikinya, dan $h_i$ yang mewakili harga anggapan bagi barangan baginya. Agen-agen 'berlanggaran' di antara satu sama lain secara rawak, dan apabila dua agen berlanggar, suatu 'urusniaga' belian berlaku: agen $j$ yang mempunyai harga anggapan yang lebih tinggi membeli dari agen $i$ yang mempunyai harga anggapan yang lebih rendah, pada harga $i$, sebanyak yang boleh, dan dengan harga anggapan $j$ menurun kepada nilai harga anggapan $i$, kecuali apabila agen $i$ tidak mempunyai barangan untuk dijual, bilamana $i$ pula membeli dari $j$ pada harga $j$ dan harga $i$ menaik. Ini ditunjukkan dalam Gambarajah 1.

Model sebegini dipilih di atas dasar mudahnya yang maksimum. Sudah tentu ia boleh diubahsuai misalnya dengan agen-agen yang berada di atas kekisi yang tegar dengan salingtindak jiranan sahaja dan samada agan boleh bergerak atau tidak, dengan amaun dibeli yang diparameterkan, dan dengan perubahan harga anggapan yang diparameterkan. Juga, perkara-perkara seperti cukai, faedah dalam duit, pengeluaran barangan, berbilang barangan, kesan teknologi, buruh, sempadan negara, dan sebagainya boleh ditambahkan dan dikaji kesannya.

---

[*] Ceramah jemputan di ITMA, UPM, Serdang, Ramadhan 1423 (8 Nov. 2002)

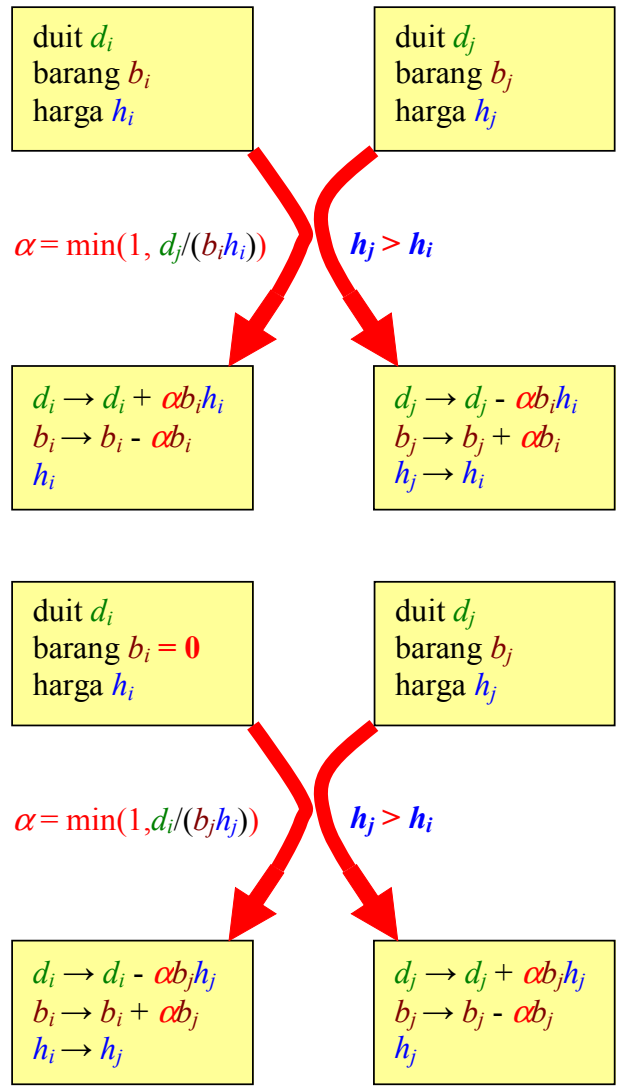

**Gambarajah 1**. Kinetik model belian

**SIMULASI**

Kita jalankan kajian model di atas menerusi simulasi komputer. Satu parameter yang boleh diubah ialah bilangan agen. Keadaan awal sistem juga boleh diubah – amaun modal yang dipunyai oleh seseorang agen, nisbah modal awal itu yang berbentuk barangan (yang lain duit), dan harga anggapan awal oleh agen bagi barangan. Kita boleh spesifikasikan nilai-nilai awal bagi parameter-parameter ini secara rawak menerusi taburan Gaussan bagi setiapnya – jadi setiapnya dicirikan oleh nilai min dan sisihan. Namun, dinamik model kita tak varian terhadap perubahan nilai jumlah modal, jadi kita boleh tetapkan nilai awalnya untuk setiap agen sebagai 1.0 tanpa hilang keumuman. Dalam kajian kita, kita tetapkan nilai modal agen awal sebagai tetap, tanpa sisihan, dan begitu juga bagi nisbah barangan awal – sesuatu nilai yang sama bagi semua agen. Ini supaya kita bermula dengan keadaan yang paling ringkas. Model kita juga takvarian terhadap nilai mutlak harga anggapan, jadi kita tetapkan nilai awalnya sebagai 1.0, tetapi dalam kes ini kita perlu sisihan supaya agen-agen mempunyai harga-harga berbeza – kita pilih sishan piawai 0.3. Kepelbagaian harga sebegini perlu dimasukkan dari awal kerana dalam model kita tiada dinamik yang membolehkan tenaga memperolehi nilai selain daripada nilai-nilai yang telah ada pada mulanya.

Simulasi dijalankan untuk 5 pusingan setiapnya, di mana satu pusingan ialah langkah masa bagi setiap agen bertembung dengan semua yang lain (secara rawak). Kita dapati sistem kita mantap setelah 1 atau 2 pusingan sahaja.

Kita kaji model belian untuk bilangan agen yang berbeza, dan juga, dengan menetapkan 100 agen, untuk nisbah barangan awal yang berbeza, yang kita semak 100 cubaan untuk setiap nilai nisbah. Kita juga cuba, sebagai perbandingan, model-model dengan dinamik yang berbeza. Akhirnya, untuk menunjukkan potensi luas yang ada dalam memperkembangkan model belian itu, kita lihat kesan faedah ke atas duit sebagai penerokaan permulaan.

## HASIL DAN PERBINCANGAN

Gambarajah 2 menunjukkan "lengkung permintaan" yang terhasil daripada model belian kita, untuk bilangan agen yang berbeza. Secara amnya, ia menunjukkan tren yang kita lihat dalam lengkung permintaan yang dijangkakan dalam ekonomi: harga rendah bila barang banyak. Kesan ketara dari bilangan agen berbeza adalah sesuatu yang boleh disiasat dengan lebih mendalam. Gambarajah 3 misalnya menunjukkan pergantungan ke atas bilangan agen ini. Perbezaan harga antara barangan sedikit dan barangan banyak lebih menonjol bagi bilangan agen banyak.

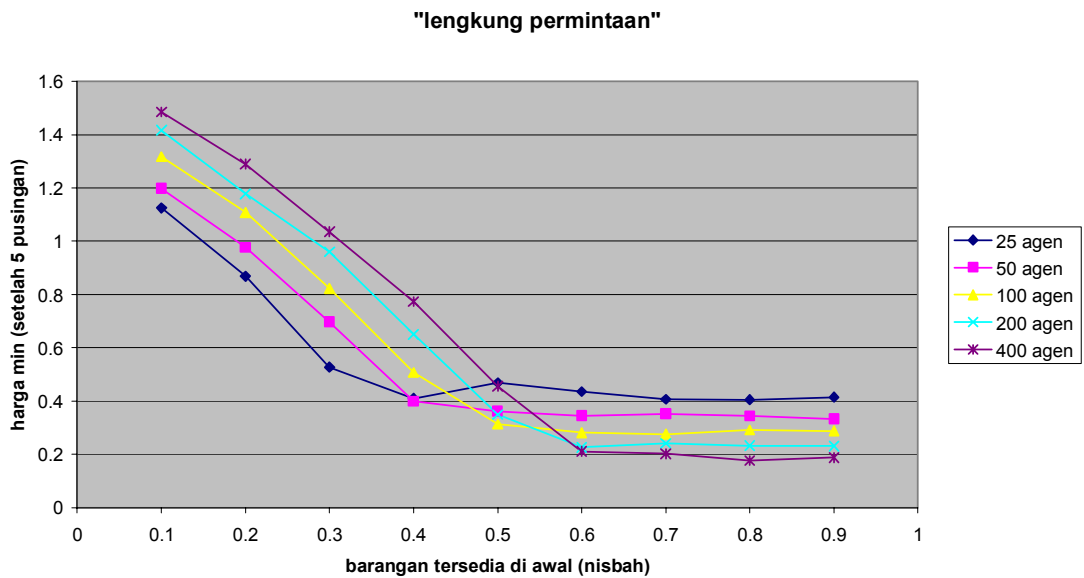

**Gambarajah 2**. Harga anggapan akhir bagi amaun barangan berbeza

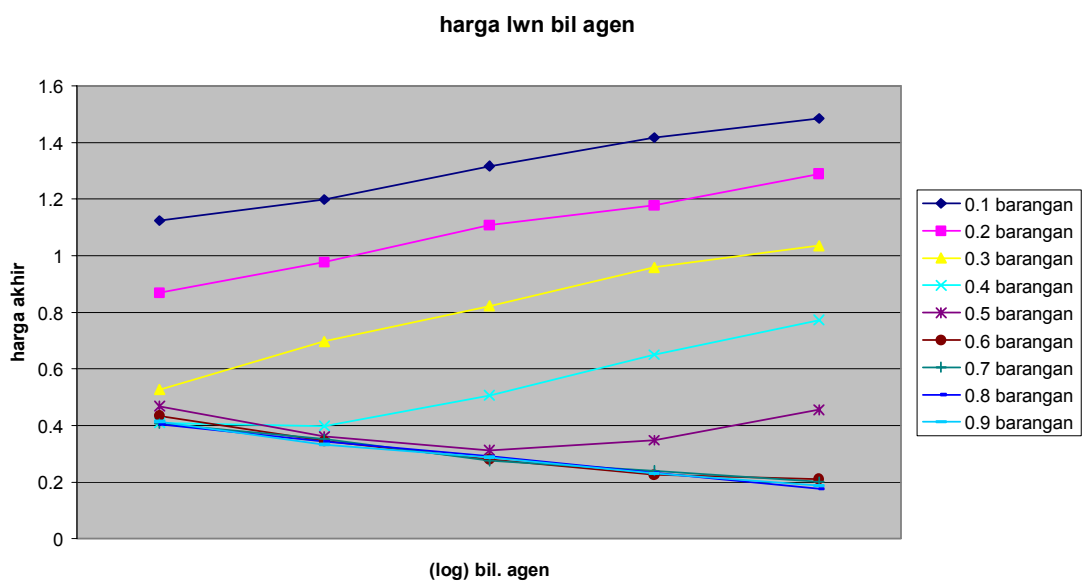

**Gambarajah 3**. Harga dengan amaun barangan berlainan, bagi bilangan agen berlainan

Kalau kita lihat sisihan dalm nilai harga anggapan akhir, di antara agen-agen, kita dapati seperti dalam Gambarajah 4. Kelihatan perubahan yang ketara pada suatu nilai tertentu, yang mungkin menandakan fasa-fasa fenomena yang berbeza.

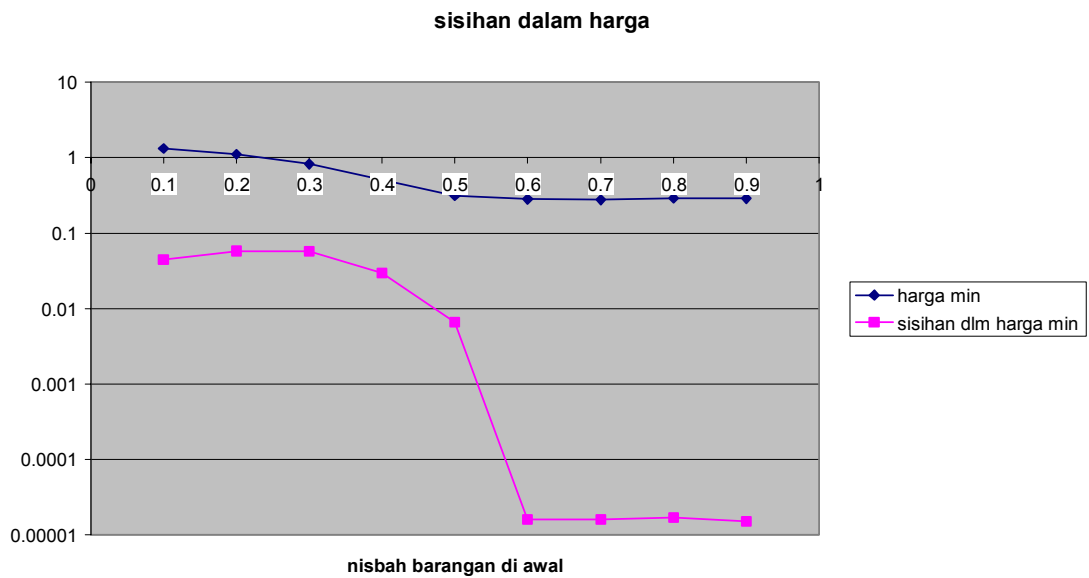

**Gambarajah 4**. Telatah sisihan harga terhadap amaun barangan

Gambarajah 5 menunjukkan jumlah kekayaan (duit campur barangan pada harga anggapan masing-masing) bagi nisbah barangan awal yang berbeza. Tren menurunnya kita jangkakan, daripada lengkung permintaan yang telah diperolehi, tapi bentuk linearnya yang agak tepat menimbulkan persoalan yang boleh dijawab oleh kajian lanjutan.

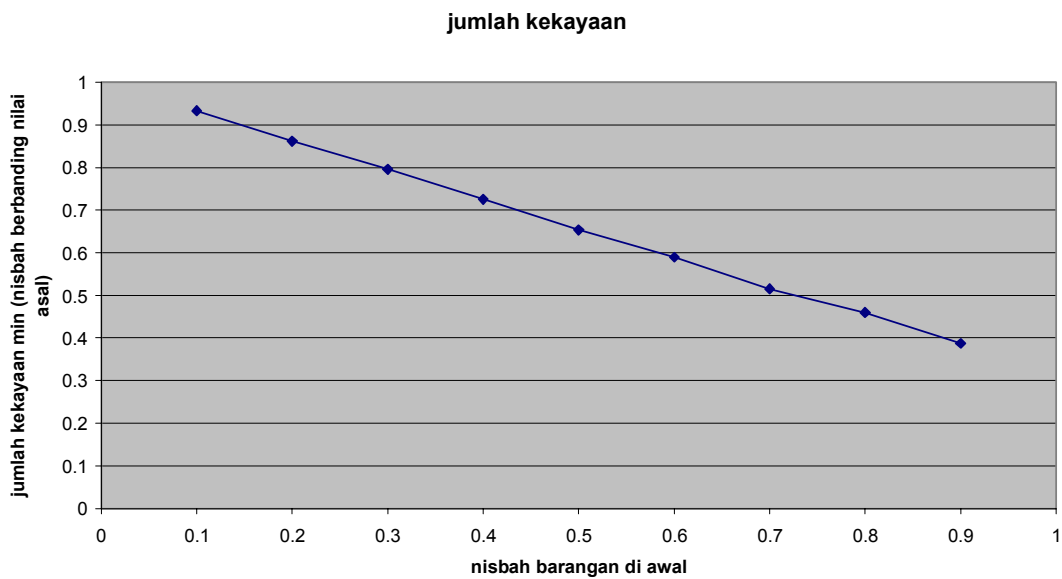

**Gambarjah 5**. Jumlah kekayaan akhir bagi amaun barangan awal berbeza

Model belian kita agak pincang kepada belian; kita boleh adakan model yang bertentangan, model 'jualan' di mana pertembungan agen menyebabkan jualan oleh $i$ kepada $j$ pada harga anggapan $j$ (harga $i$ naik kepada nilai harga $j$, yang tak berubah), kecuali bila $j$ kehabisan duit bilamana $j$ menjual kepada $i$ dengan harga $i$ (harga $j$ turun ke harga $i$). Adakah kita akan lihat fenomena lengkung permintaan dalam model jualan ini seperti yang kita lihat dalam model belian? Jawapannya tertera dalam Gambarajah 6 – tiada pergantungan harga anggapan akhir kepada nisbah amaun barangan. Harga keseluruhan

meningkat – tapi kepada harga yang sama tidak kira amaun barangan. Pemahaman ini perlukan kajian lanjut, begitu juga dengan pemahaman kenapa ada perbezaan di antara model belian dan model jualan. Mungkinkah ini berkaitan dengan ketaksimetrian untung-rugi yang telah dikesan di tempat lain [2] ? Pemahaman tentang bentuk linear dan meningkat kekayaan akhir jumlah bagi model jualan ini, seperti dalam Gambarajah 7, perlu juga dicari.

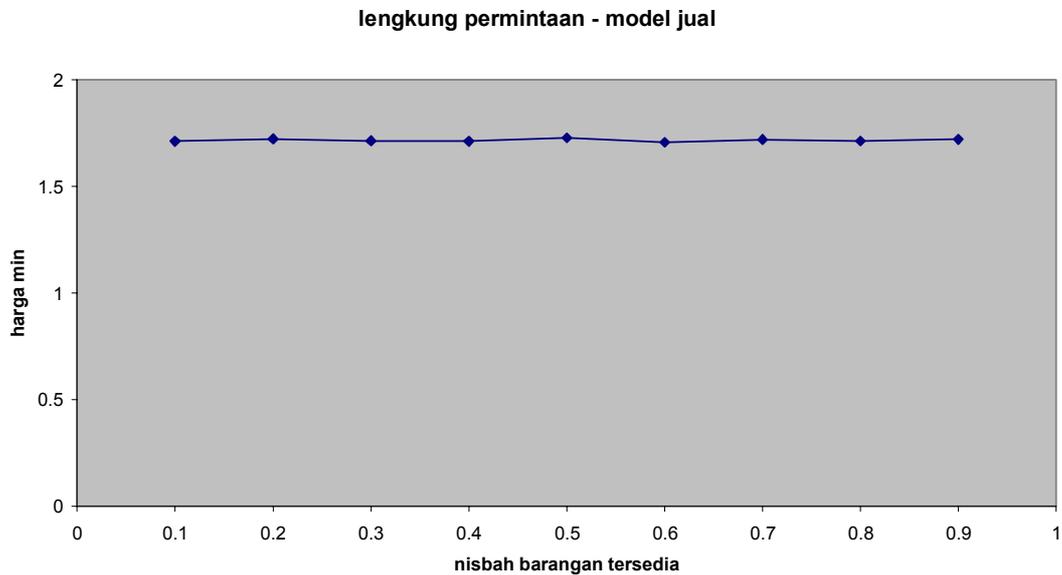

**Gambarajah 6**. Lengkung permintaan bagi model jualan

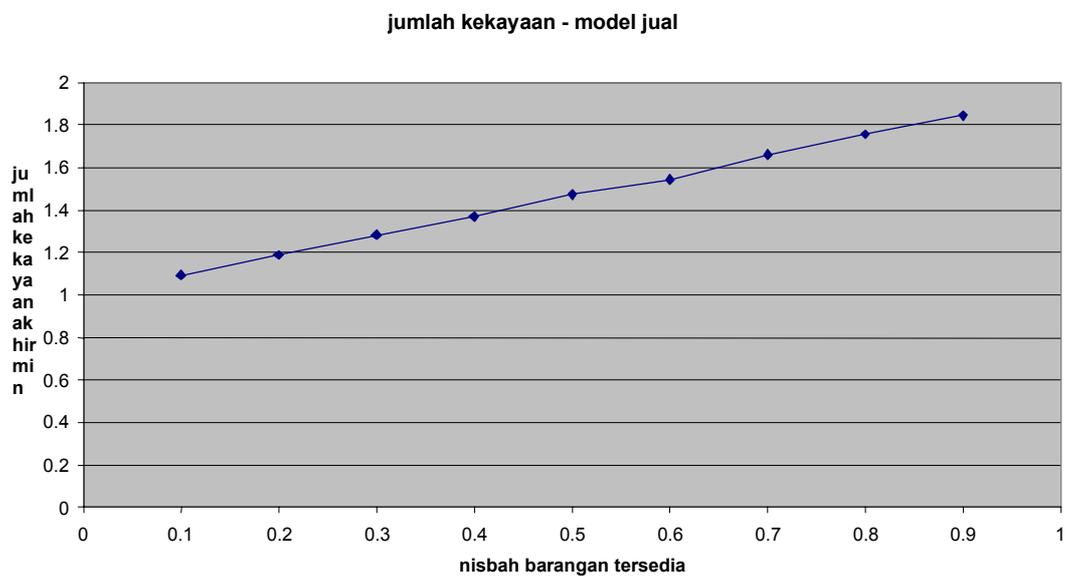

**Gambarajah 7**. Jumlah kekayaan akhir bagi amaun barang berlainan untuk model jualan

Bagi model yang bersimetri penuh, di mana siapa yang menjual atau membeli (dan harga yang berubah) tidak bergantung kepada harga anggapan agen, tetapi secara rawak, kita jangkakan harga tak berubah secara purata, seperti yang didapati dan ditunjukkan dalam Gambarajah 8. Gambarajah 9 menunjukkan kekayaan jumlah akhir yang diperolehi bagi model simetri ini, dan, kecualikan sedikit tren menaik terhadap amaun barangan, adalah lebih-kurang tetap pada nilai asal secara purata.

Jelaslah bahawa lengkung permintaan yang diperolehi dalam model belian yang asal itu agak bitara, dan tidak ditunjukkan dalam model akasnya, iaitu model jualan, dan tidak juga dalam model bersimetri. Yang menarik ialah ia timbul daripada dinamik agen yang sangat ringkas.

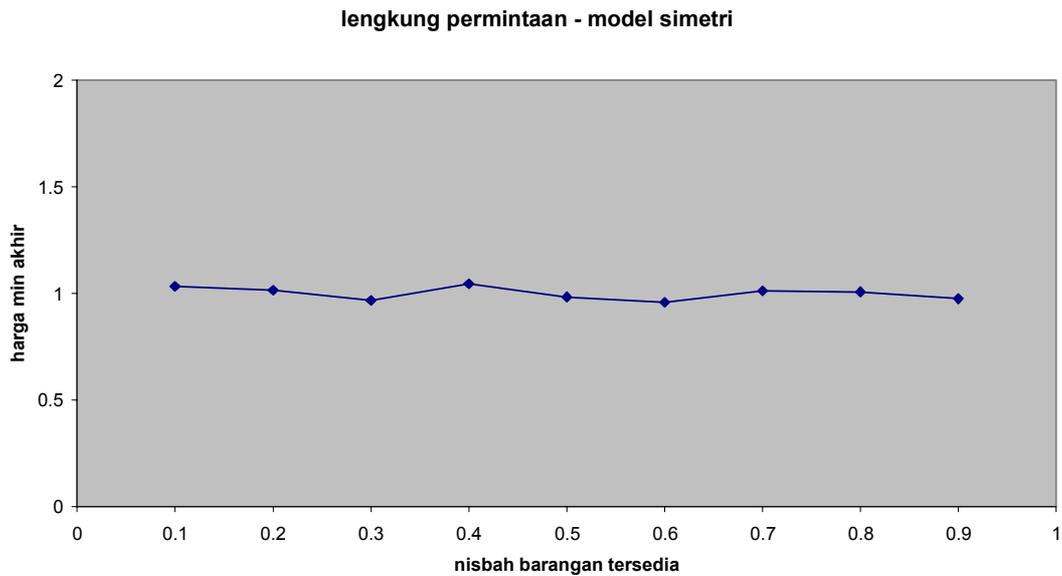

**Gambarajah 8**. Lengkung permintaan bagi model bersimetri

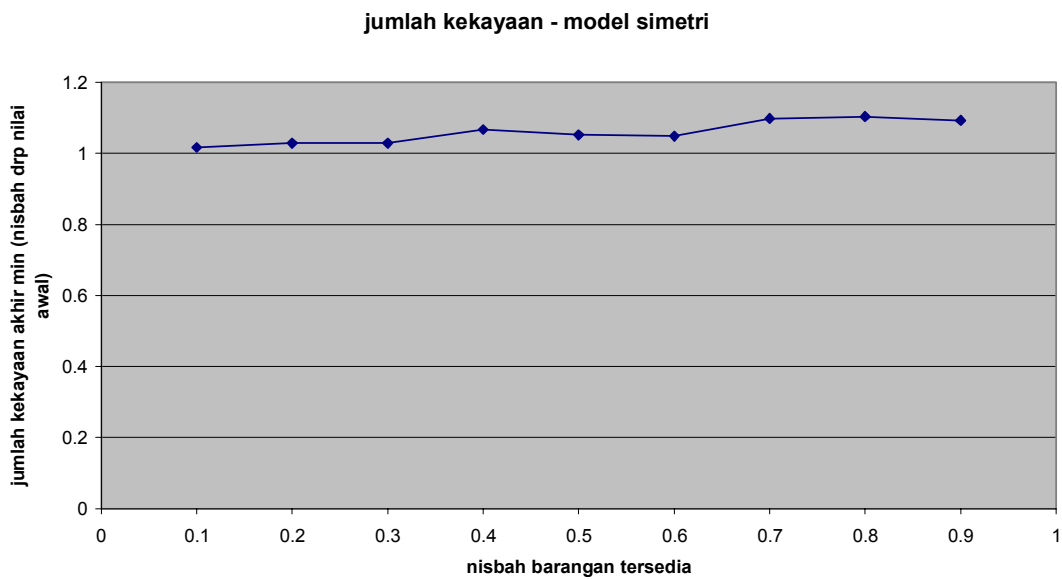

**Gambarajah 9**. Jumlah kekayaan akhir bagi amaun barangan berbeza untuk model bersimetri

Model belian asas kita sudah tentu boleh diperluaskan dalam pelbagai aspek. Sebagai penerokaan awal, kita lihat kesan faedah. Gambarajah 10 menunjukkan bagaimana harga anggapan min berubah terhadap masa dengan duit diberikan faedah 0.1 (10 %) setiap pusingan atau langkah masa. Bagi keadaan di mana barangan banyak, kesan faedah ini tidak sempat dirasai kerana harga telah menjunam beku ke nilai-nilai rendah tertentu. Untuk amaun barangan awal yang nisbahnya rendah daripada nilai genting tertentu, harga terapung-apung naik secara perlahan. Pemahaman mendalam mengharuskan kajian lanjut.

**RUMUSAN**

Kita telah memperkenalkan satu model dengan dinamik ringkas yang telah menimbulkan telatah seakan ekonomi dalam kajian awal ini. Model yang lebih pincang kepada belian lebih mirip kenyataan daripada model yang pincang jualan atau model bersimetri. Banyak fenomena yang menarik diperhatikan yang memanggil kajian lanjutan. Perluasan model ini dalam pelbagai aspek juga menjanjikan penerokaan seterusnya yang menarik.

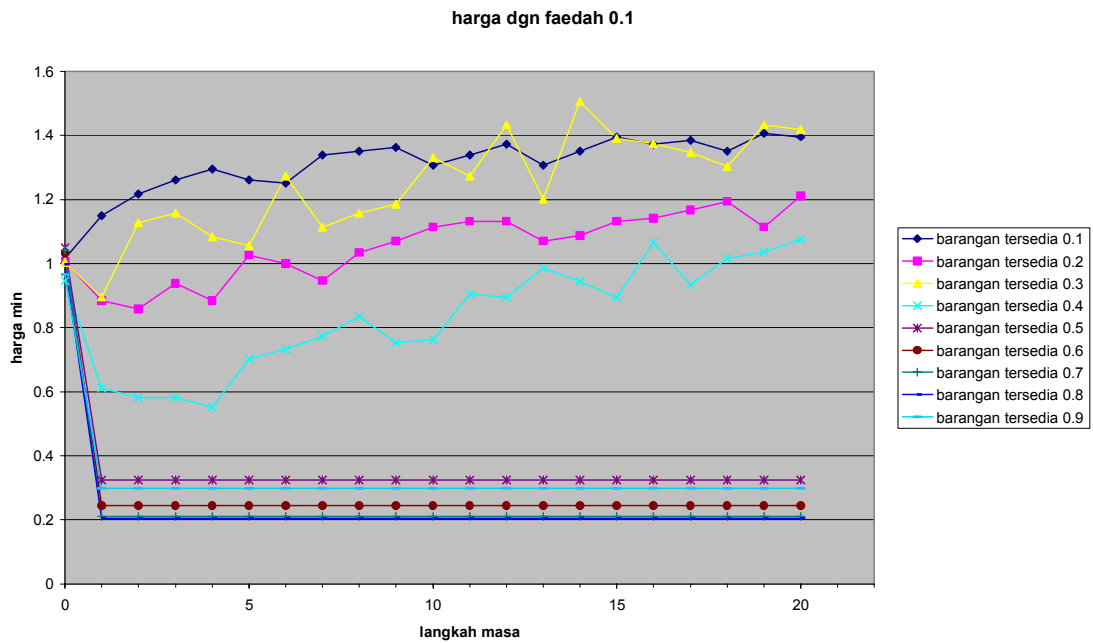

**Gambarajah 10**. Kesan faedah ke atas harga anggapan, bagi amaun barangan berlainan